\documentclass[10pt,english]{article}
\usepackage{amsfonts}
\usepackage{amssymb,amsthm,amsmath}
\usepackage{graphicx}
\usepackage{graphics}
\usepackage{subfigure}
\usepackage{url}

\newtheorem{theorem}{Theorem}[section]

\newtheorem{proposition}[theorem]{Proposition}
\newtheorem{remark}[theorem]{Remark}

\begin{document}

\title{Pricing  Exchange Options under  Stochastic Correlation }
\author{Enrique Villamor and Pablo Olivares}

%\textbf{Department of Mathematics. Florida International University. Miami, FL 33199, USA and Department of Mathematics. Ryerson University. Toronto, Canada}
\maketitle

\begin{abstract}
In this paper we study the pricing of exchange options when underlying assets  have stochastic volatility and stochastic correlation. An approximation using a closed-form approximation based on a Taylor expansion of the conditional price is proposed. Numerical results are illustrated for exchanges between WTI and Brent type oil prices.
\end{abstract}

%\date{Today}

\section{Introduction}
In this paper we study the pricing of exchange options when the underlying assets have stochastic volatility and correlation. Its main contribution is the proposal of a approximated closed-form formula under this framework. \\
The exchange of two assets is used to  hedge against the changes in price of  underling assets by betting on the difference between both.\\
 The price of these instruments has been first considered in \cite{marg} under a standard bivariate Black-Scholes model, where a closed-form formula for the pricing is provided.  The results have been extended in \cite{chea2,chea} to   the case of a  jump-diffusion model, while  in \cite{exchir} it has been considered the pricing of the derivative under  stochastic interest rates.\\
   It is well known that constant correlation and  volatilities assumed in the context of a Black-Scholes model  are  not supported by empirical evidence. In the seminal paper of Heston, see \cite{Hes}, the pricing of option contracts under  stochastic volatility is studied. The idea is extended to stochastic correlation in \cite{AEO12}, while still considering constant volatilities.\\
   As an alternative view to correlation, models for the covariance process have been proposed. The pricing of exchanges under stochastic covariance is adopted  in Olivares and Villamor(2018), see \cite{Olvillamor}. See for example \cite{Fonseca} for the  Wishart model and \cite{pigor} for an Ornstein-Uhlenbeck Levy type model.\\
   We consider a bivariate continuous-time GARCH process to  model the correlation combined with a pricing method based on a Taylor expansion of the conditional Margrabe price.  Continuous-time GARCH processes as limits of the embedded  discrete-time counterpart have been proposed, for example, in \cite{drost} or \cite{brock}. See also \cite{hull12}. \\
   The organization of the paper is the following:\\
In section 2 we introduce the model, discuss the approximated pricing formula and compute the first and second order moment of the underlying assets, their volatilities and their correlations, whose proofs are deferred to the appendix. In section 3 we discuss the numerical results for the pricing of exchange options between WTI and Brent type oil futures. Finaly, we present the conclusions. 
 \section{Pricing exchange options in models with stochastic correlation }
Let  $(\Omega ,\mathcal{F}, (\mathcal{F}_{t})_{t \geq 0}, P)$ be a filtered probability space. We denote  by $\mathcal{Q}$ a risk-neutral  equivalent martingale  measure(EMM) and $E_{\mathcal{Q}}$ the expected value with respect to the measure $\mathcal{Q}$. For a process $(X_t)_{t \geq 0}$, the integrated process associated with it is denoted by $(X^+_t)_{t \geq 0}$ and defined as:
\begin{equation*}
  X^+_t= \int_0^t X_s\;ds
\end{equation*}
 The functions $f_{X}(x)$ and $f_{X/Y}(x/y)$ are respectively the probability density function (p.d.f.)  of the random vector $X$ and the conditional p.d.f. of the random vector $X$ on the random vector $Y$.\\
 A  two-dimensional adapted stochastic process $(S_t)_{t \geq 0}=(S^{(1)}_t,S^{(2)}_t)_{t \geq 0}$, where their components are prices of certain underlying assets, is defined on the filtered probability space above.\\
We assume that the process of  prices  has a dynamic under  $\mathcal{Q}$  given by:
\begin{eqnarray}\label{modpri1}
dS^{(1)}_t &=& r\,S^{(1)}_t\, dt +\sigma_t^{(1)}\,S^{(1)}_t\, dZ_t^{(1)}\\ \label{modpri2}
dS^{(2)}_t &=&r\,S^{(2)}_t\, dt +\sigma_t^{(2)}\,\sqrt{1-\rho_t^{(2)}}\,S^{(2)}_t\, dZ_t^{(2)}+\sigma_t^{(2)}\,\rho_t\,S^{(2)}_t\,dZ_t^{(1)}
\end{eqnarray}
where the $(\sigma_t)_{t \geq 0}=(\sigma_t^{(1)},\sigma_t^{(2)})_{t \geq 0}$ is the volatility process  and  $\rho_{t}$ is the instantaneous correlation coefficient, which in our models are going to be stochastic. \\
 The payoff of a European  exchange option, with maturity at time $T>0$ is:
\begin{equation}\label{eq:payoff}
      h(S_T)=(cS^{(1)}_T-m S^{(2)}_T)_+
\end{equation}
where $m$ is the number of assets of type two exchanged against $c$ assets of type one. To simplify we assume $c=m=1$.\\
The volatilities are modeled as an Ornstein-Ulenbeck processes:
\begin{equation}\label{eq:vol}
  d\sigma_t^{(j)}=-\alpha_j \sigma_t^{(1)}+ \beta_j  dW_t^{(j)},\; j=1,2
\end{equation}
The Brownian motions $(W_t^{(1)})_{t \geq 0}$ and $(W_t^{(2)})_{t \geq 0}$ have instantaneous correlation $\rho_V$.\\
By Ito formula:
 \begin{eqnarray}\label{eq: svol1}
    dV^{(j)}_t&=& c_1(V^{(j)}_L-V^{(j)}_t)\,dt +\xi_j  \sigma^{(j)}_t\, dW_t^{(j)},\; j=1,2 \\ \label{eq: corr}
d \rho_t &=& \bar\gamma\,(\bar\Gamma_L- \,\rho_t)\,dt+\bar\alpha  \,\sqrt{1- \rho_t^2 }\,d\bar W_t
 \end{eqnarray}
 where $V_t=(V^{(1)}_t , V^{(2)}_t)_{t \geq 0}$, with $V^{(j)}_t=(\sigma_t^{(j)})^2, j=1,2$ is the process of squared volatilities. The parameters $V_L=(V^{1}_L, V^{2}_L)$  and $c_j>0$ are respectively the mean-reverting level and rate  of the squared volatility processes while $\bar\Gamma_L$ and $\bar\gamma$ play a similar role in the correlation process. \\
 The  two components of the Brownian  motion $(Z_t)_{t \geq 0}=(Z_t^{(1)},\,Z_t^{(2)})_{t \geq 0}$  are  assumed to be independent of the second set of Brownian motions $(W_t)_{t \geq 0}=(W_t^{(1)},\, W_t^{(2)})_{t \geq 0}$ and $\bar W_t$.
 \begin{remark}\label{rem:par}
 Parameters in  models (\ref{eq:vol}) and (\ref{eq: svol1}) are related by $c_j=2 \alpha_j$, $V^j_L=\frac{\beta^2_j}{2 \alpha_j}$ and $\xi_j=2 \beta_j$.
 \end{remark}
   % See appendix A for the building of the continuous-time analogous of a GARCH(1,1) model  for the variance and correlation processes.\\
Next, we find an expression for the price of the exchange contract. Notice that  the price of this contract at time $t$, $0 \leq t \leq T$  with maturity at $T$   is given by:
\begin{equation}\label{eq:pri}
  C_t=e^{-r(T-t)}E_{\mathcal{Q}}[ h(S_T)]
\end{equation}
Its terminal value is $C_T=h(S_T)$.\\
The price of the exchange contract at time $t$, $t< T$, depends on the behavior of the  processes $( V_s, \rho_s)_{ t \leq s \leq T}$  described by equations (\ref{eq: svol1})-(\ref{eq: corr}) and integrated on the interval $[t,T]$. It depends also on the spot prices, volatilities and correlation at time $t$. For simplicity in the notations we explicitly drop this last dependence. For the same reason, we analyze only the case $t=0$. \\
 Hence:
\begin{eqnarray}\nonumber
  C_0 &=&  e^{-r T}\,\,\int_{\mathbb{R}^5}h(x)f_{S_T, V^+_T, \rho^+_T}(x) \;dx\\ \nonumber
   &=&  e^{-r T}\,\,\int_{\mathbb{R}^3}\left[\int_{\mathbb{R}^2}C_T(x', x'')f_{S_T/ V^+_T, \rho^+_T}(x'/x'') \right]f_{ V^+_T, \rho^+_T}(x'')\;dx'' \\ \label{eq:partpri}
   &=&  \int_{\mathbb{R}^3}C_M(x'')f_{V^+_T, \rho^+_T}(x'')\;dx''
\end{eqnarray}
 where $x=(x', x'') \in \mathbb{R}^5$. \\
The function $C_M(x'')=e^{-r(T-t)}\int_{\mathbb{R}^2}C_T( x',x'')f_{S_T/ V^+_T, \rho^+_T}(x'/x'')\;dx'$ is the Margrabe price conditionally on $(V^+_T, \rho^+_T)=x''$. After conditioning it equals the Margrabe price, see \cite{marg}. A closed-form for the latter is given by:
\begin{equation}\label{eq:pricemag}
    C_{M}(V^+_T, \rho^+_T)=e^{-r T}S^{(1)}_t N(d_1(v^+_T))-e^{-rT}S^{(2)}_t N(d_2(v^+_T))
\end{equation}
with:
\begin{eqnarray*}
    d_1(v^+_T)&=&\frac{\log \left(\frac{ S^{(1)}_t}{ S^{(2)}_t} \right)+\frac{1}{2}v^+_T }{\sqrt{v^+_T} }\\
    d_2(v^+_T)&=&\frac{\log \left(\frac{ S^{(1)}_t}{ S^{(2)}_t} \right)-\frac{1}{2}v^+_T }{\sqrt{v^+_T} }=d_1(v^+_T)-\sqrt{v^+_T}
\end{eqnarray*}
where:
\begin{equation*}
  v^+_T=V^{1,+}_T+V^{2,+}_T-2 \sqrt{V^{1,+}_T V^{2,+}_T} \rho^+_T
\end{equation*}
and $(V^+_t)_{t \geq 0}=(V^{1,+}_t, V^{2,+}_t)_{t \geq 0}$.\\
Next,  to approximate the price in equation (\ref{eq:pri}) we consider a second order Taylor expansion of the conditional Margrabe price $C_{M}(x), x \in \mathbb{R}^3$  around the average values  given by $x_0=(E_{\mathcal{Q}}(V^{1,+}_T), E_{\mathcal{Q}}(V^{2,+}_T), E_{\mathcal{Q}}(\rho^+_T))$. It leads to:
\begin{eqnarray}\nonumber
  \hat{C}_M(x) &=&  C_M(x_0)+\frac{\partial  C_M(x_0)}{\partial x_1}(x_1-x_{0,1})+\frac{\partial C_M(x_0)}{\partial x_2}(x_2-x_{0,2})+\frac{\partial  C_M(x_0)}{\partial x_3}(x_{0,3}-x_0)\\ \nonumber
  &+& \frac{1}{2}\frac{\partial^2  C_M(x_0)}{\partial x^2_1}(x_1-x_{0,1})^2 + \frac{1}{2}\frac{\partial^2  C_M(x_0)}{\partial x^2_2}(x_2-x_{0,2})^2 \\ \nonumber
 &+& \frac{1}{2} \frac{\partial^2  C_M(x_0)}{\partial x^2_{3}}(x_3-x_{0,3})^2 + \frac{\partial^2  C_M(x_0)}{\partial x_1 x_2}(x_1-x_{0,1})(x_2-x_{0,2})\\ \nonumber
 &+& \frac{\partial^2  C_M(x_0)}{\partial x_1 x_3}(x_1-x_{0,1})(x_3-x_{0,3})+ \frac{\partial^2  C_M}{\partial x_2 x_3}(x_0)(x_2-x_{0,2})(x_3-x_{0,3})\\ \label{eq:taylor}
 &&
\end{eqnarray}
Combining equations (\ref{eq:partpri}) and (\ref{eq:taylor}) we have the price $C_0$  is approximated by:
\begin{eqnarray}\nonumber
  \hat{C}_0 &=&  C_M(x_0)+ \frac{1}{2}\frac{\partial^2  C_M(x_0)}{\partial x^2_1}Var_{\mathcal{Q}}(V^{1,+}_T) + \frac{1}{2}\frac{\partial^2  C_M(x_0)}{\partial x^2_2}Var_{\mathcal{Q}}(V^{2,+}_T) \\ \nonumber
 &+& \frac{1}{2} \frac{\partial^2  C_M(x_0)}{\partial x^2_{3}}Var_{\mathcal{Q}}(\rho^+_T) + \frac{\partial^2  C_M(x_0)}{\partial x_1 x_2}cov_{\mathcal{Q}}(V^{1,+}_T, V^{2,+}_T)\\  \label{eq:price}
\end{eqnarray}
  Notice that the Margrabe price $C_M(x) \in C^{\infty}(\mathbb{R}^3)$ except in a set of zero Lebesgue measure.\\
  We substitute equation (\ref{eq:taylor}) into (\ref{eq:partpri}). Noticing that:
\begin{eqnarray*}
\int_{\mathbb{R}^3} (x_1-x_{0,1})f_{V^+_T, \rho^+_T}(x)\;dx &=& \int_{\mathbb{R}} (x_1-x_{0,1})\left[\int_{\mathbb{R}^2}f_{V^+_T, \rho^+_T}(x)\;dx_2\;x_3 \right]\;x_1\\
&=& \int_{\mathbb{R}} (x_1-x_{0,1})f_{V^{1,+}_T}(x_1) dx_1=E_{\mathcal{Q}}(V^{1,+}_T-E_{\mathcal{Q}}(V^{1,+}_T))=0\\
\int_{\mathbb{R}^3} (x_2-x_{0,2})f_{V^+_T, \rho^+_T}(x)\;dx &=& \int_{\mathbb{R}} (x_2-x_{0,2})\left [\int_{\mathbb{R}^2}f_{V^+_T, \rho^+_T}(x)\;dx_1\;x_3 \right]\;x_2\\
&=& \int_{\mathbb{R}} (x_2-x_{0,2})f_{V^{2,+}_T}(x_2) dx_2=E_{\mathcal{Q}}(V^{2,+}_T-E_{\mathcal{Q}}(V^{2,+}_T))=0\\
\int_{\mathbb{R}^3} (x_3-x_{0,3})f_{V^+_T, \rho^+_T}(x)\;dx &=& \int_{\mathbb{R}} (x_3-x_{0,3})\left [\int_{\mathbb{R}^2}f_{V^+_T, \rho^+_T}(x)\;dx_1\;x_2 \right]\;x_3\\
&=& \int_{\mathbb{R}} (x_3-x_{0,3})f_{\rho^{+}_T}(x_3) dx_3=E_{\mathcal{Q}}(\rho^{+}_T-E_{\mathcal{Q}}(\rho^{+}_T))=0\\
\end{eqnarray*}
Hence, we have equation (\ref{eq:price}).
  \begin{remark}
 Sensitivities with respect to the parameters in the contract can be computed in a similar way. For example, an approximation of   the deltas in the exchange contract are obtaining by differentiating equation (\ref{eq:price}) with respect to the price of the underlying assets.
 \end{remark}
 Computing derivatives of the Margrabe price, given by  equation (\ref{eq:pricemag}), with respect to the volatilities and correlation is straightforward. This aspect is addressed in appendix B.\\
In order to estimate the option pricing function above we need to compute the moments of $(V^{1,+}_T, V^{+,2}_T,\rho^+_T)$. To this end we introduce the following notations:
\begin{eqnarray*}
  mr_j(t)&=& E[\rho^j_t],\; mr^+_j(t)=E[(\rho^+_t)^j],\; j=1,2\\
  mv_{j,k}(t)&=& E[(V^{(k)}_t)^j],\; mv^+_{j,k}(t)=E[(V^{k,+}_t)^j]\; j,k=1,2\\
  mv_{12}(t)&=& E[V^{(1)}_t V^{(2)}_t],\; mv^+_{12}(t)= E[V^{1,+}_t V^{2,+}_t]
\end{eqnarray*}
Results are given in the propositions below, while proofs are deferred to appendix A.
\begin{proposition}\label{prop:momentsrho}
Let the correlation process $( \rho_t)_{t \geq 0}$  satisfy equation (\ref{eq: corr}). Then:
\begin{eqnarray}\label{eq:exprho}
  E_{\mathcal{Q}}(\rho^+_t)  &=& \bar \Gamma_L t+\left(\frac{\rho_0-\bar \Gamma_L}{\bar\gamma}\right)(1-e^{-\bar\gamma t}) \\ \nonumber
  Var_{\mathcal{Q}}(\rho^+_t) &=& b_0+ \left(\frac{\rho_0-\bar \Gamma_L}{\bar\gamma}\right)^2 +\left(b_1+2\bar \Gamma_L \left(\frac{\rho_0-\bar \Gamma_L}{\bar\gamma}\right)\right)t\\ \nonumber
  &+& (b_2+ \bar\Gamma^2_L) t^2+ \left(b_3- 2 \bar\Gamma_L \left(\frac{\rho_0-\bar \Gamma_L}{\bar\gamma}\right) \right)t e^{-\bar\gamma t}+b_4 e^{-(2\bar\gamma+\bar\alpha^2)t}\\ \nonumber
  &-& \left(b_0+b_4+2 \left(\frac{\rho_0-\bar \Gamma_L}{\bar\gamma}\right)^2 \right) e^{-\bar\gamma t}+ \left(\frac{\rho_0-\bar\Gamma_L}{\bar\gamma}\right)^2 e^{-2 \bar\gamma t}\\ \label{eq:varrho}
  &&
 \end{eqnarray}
where:
\begin{eqnarray*}
a_1 &=& \frac{2 \bar\gamma \bar \Gamma^2_L+ \bar\alpha^2 }{2\bar\gamma+\bar\alpha^2},    a_2 = \frac{2 \bar\gamma \bar \Gamma_L(\rho_0-\bar \Gamma_L)}{\bar\gamma+\bar\alpha^2}\\
b_0 &=&  \frac{1}{\bar\gamma^2} \left( -a_1+\rho^2_0-2 \bar \Gamma_L+\frac{2}{\bar\gamma^2} \right.\\
&-& \left. \frac{\bar \alpha^2}{\bar\gamma^2}\left(1+\frac{a_1}{\bar\gamma}+ a_2\right)-\frac{\alpha^2 (\rho^2_0-a_1-a_2)}{ \bar\gamma(2 \bar\gamma+\bar\alpha^2)} \right)\\
 b_1 &=& \frac{1}{\bar\gamma} \left( -a_2+ 2 \bar \Gamma_L- \frac{2}{\bar\gamma^2}+ \frac{\bar\alpha^2}{\bar\gamma}- \frac{a_1 \bar\alpha^2}{\bar\gamma^2} \right) \\
 b_2 &=& 1, \; b_3 = \frac{a_2 \bar\alpha^2}{\bar\gamma^3}  \\
b_4 &=& -\frac{\bar\alpha^2}{\bar\gamma^2}\frac{(\rho^2_0-a_1-a_2)}{(2 \bar\gamma+\bar\alpha^2)(\bar\gamma+\bar\alpha^2)}
\end{eqnarray*}
\end{proposition}
Second order  moments and covariance of the integrated squared volatility are given in the  propositions above:
\begin{proposition}\label{prop:momentsv}
 Let the process $( V_t)_{t \geq 0}$  satisfy equations (\ref{eq: svol1})-(\ref{eq: vol2}). Then:
 \begin{eqnarray}\label{eq:firstmomvol}
mv^+_{1,j}(t) &=&  V^{(j)}_L t +\frac{V^{(j)}_0-V^{(j)}_L}{c_j}(1-e^{-c_jt}) \\ \nonumber
mv^+_{2,j}(t) &=& P_1(t)+c e^{-c_j t}+g_0 e^{-2c_j t}+g_1 e^{-3c_j t}+g_2 t e^{-c_j t}\\
 Var_{\mathcal{Q}}[V_t^{+,j}]&=& mv^+_{2,j}(t)-[mv^+_{1,j}(t)]^2
   \end{eqnarray}
 with:
  \begin{eqnarray*}
P_1(t)&=&\frac{1}{c^3_j}(V^{(j)}_L)^2 t^2+\frac{1}{c^3_j} \left((2-\frac{V^{(j)}_L}{c_j})V^{(j)}_L+ \xi^2_j  V^{(j)}_L\right)t\\
&+& \frac{1}{c^3_j} \left( (V^{(j)}_0)^2+\frac{2(V^{(j)}_L)^2}{c^2_j}- \frac{\xi^2_j  V^{(j)}_L}{c_j}\right)\\
c &=& \frac{1}{c^3_j}(\frac{1}{c_j}\xi^2_j  V^{(j)}_L+d_0+\frac{d_1}{2}-\frac{2(V^{(j)}_L)^2 }{c^2_j})\\
g_0 &=& - \frac{1}{c^3_j}[d_0+(V^{(j)}_0)^2-d_1]\\
g_1 &=& - \frac{1}{c^3_j},g_2 = \frac{1}{c^3_j} \left[\xi^2_j  ( V^{(j)}_0-V^{(j)}_L) \right]
 \end{eqnarray*}
 \end{proposition}
 \begin{proposition}\label{eq:covvol}
 Let the process $( V_t)_{t \geq 0}$  satisfy equations (\ref{eq: svol1})-(\ref{eq: vol2}). Then:
 \begin{eqnarray}\nonumber
  cov( V^{(1)}_t,  V^{(2)}_t) &=&  mv^+_{12}- mv^+_{1,1}(t)mv^+_{1,2}(t)
 \end{eqnarray}
 where:
\begin{eqnarray}\nonumber
 mv^+_{12}(t) &=&  E_{\mathcal{Q}}[V_t^{+,1}V_t^{+,2}]= \frac{1}{c_1 c_2} \left[ P_3(t)- (V^{(1)}_0+ c_1 V^{(1)}_L t) mv_{1,2}(t) - (V^{(2)}_0+c_2 V^{(2)}_L t )mv_{1,1}(t) \right.\\ \nonumber
   &+& \left. ms_{12}(t) -\xi_1 \xi_2  \rho_V e^{-c_1 t} B_1(t) -  \xi_1 \xi_2 \rho_V e^{-c_2 t}B_2(t) + \xi_1 \xi_2  \rho_V A(t) \right]
 \end{eqnarray}
   where:
    \begin{eqnarray*}
    P_3(t)&=& V^{(1)}_0 V^{(2)}_0+ c_2 V^{(1)}_0  V^{(2)}_L t+ c_1 V^{(2)}_0 V^{(1)}_L t  +c_1 c_2 V^{(1)}_L V^{(2)}_L t^2 \\
    A(t) &=& \frac{\xi_1 \xi_2 \rho_V}{2(c_1+c_2)}\left( t- \frac{2}{c_1+c_2}(1-e^{-\frac{1}{2}(c_1+c_2)t}) \right)+\frac{2(\sigma_0^{(1)} \sigma_0^{(2)})}{c_1+c_2}((1-e^{-\frac{1}{2}(c_1+c_2)t}))\\
    B_j(t) &=& \frac{\xi_1 \xi_2 \rho_V}{2(c_1+c_2)}\left(\frac{1}{c_j}(e^{c_j t}-1) - \frac{2(-1)^j}{c_2-c_1}(e^{\frac{1}{2}(-1)^j(c_2-c_1)t}-1) \right)\\
     &+& \sigma_0^{(1)} \sigma_0^{(2)}(\frac{2(-1)^j}{c_2-c_1}(e^{\frac{1}{2}(-1)^j(c_2-c_1)t}-1))
   \end{eqnarray*}
   The functions $m^+_{1,j}(t)$ are given by equation (\ref{eq:firstmomvol}) while:
   \begin{eqnarray*}
  ms_{12}(t)  &=&  \frac{\xi_1 \xi_2 \rho_V}{2(c_1+c_2}\left( 1-\exp(-\frac{1}{2}(c_1+c_2)t) \right)+\sigma_0^{(1)} \sigma_0^{(2)}\exp(-\frac{1}{2}(c_1+c_2)t)\\
  m_{1,j}(t) &=& V^{(j)}_L +(V^{(j)}_0-V^{(j)}_L)e^{-c_jt}
   \end{eqnarray*}
 \end{proposition}
\begin{figure}[htb!]
\begin{center}
\subfigure[]{
\resizebox*{7cm}{!}{\includegraphics{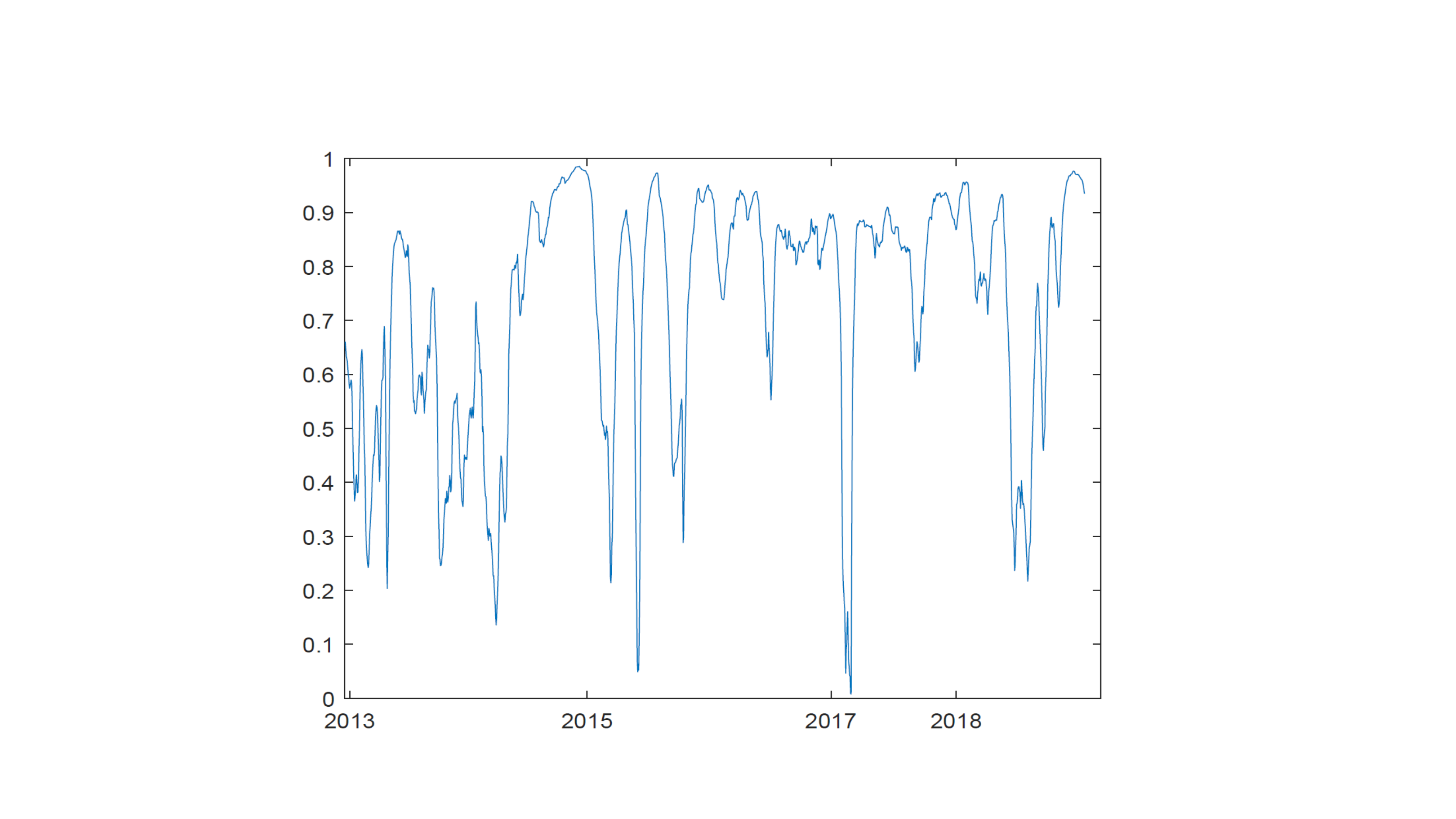}}}%\hspace{5pt}
\subfigure[]{
\resizebox*{7cm}{!}{\includegraphics{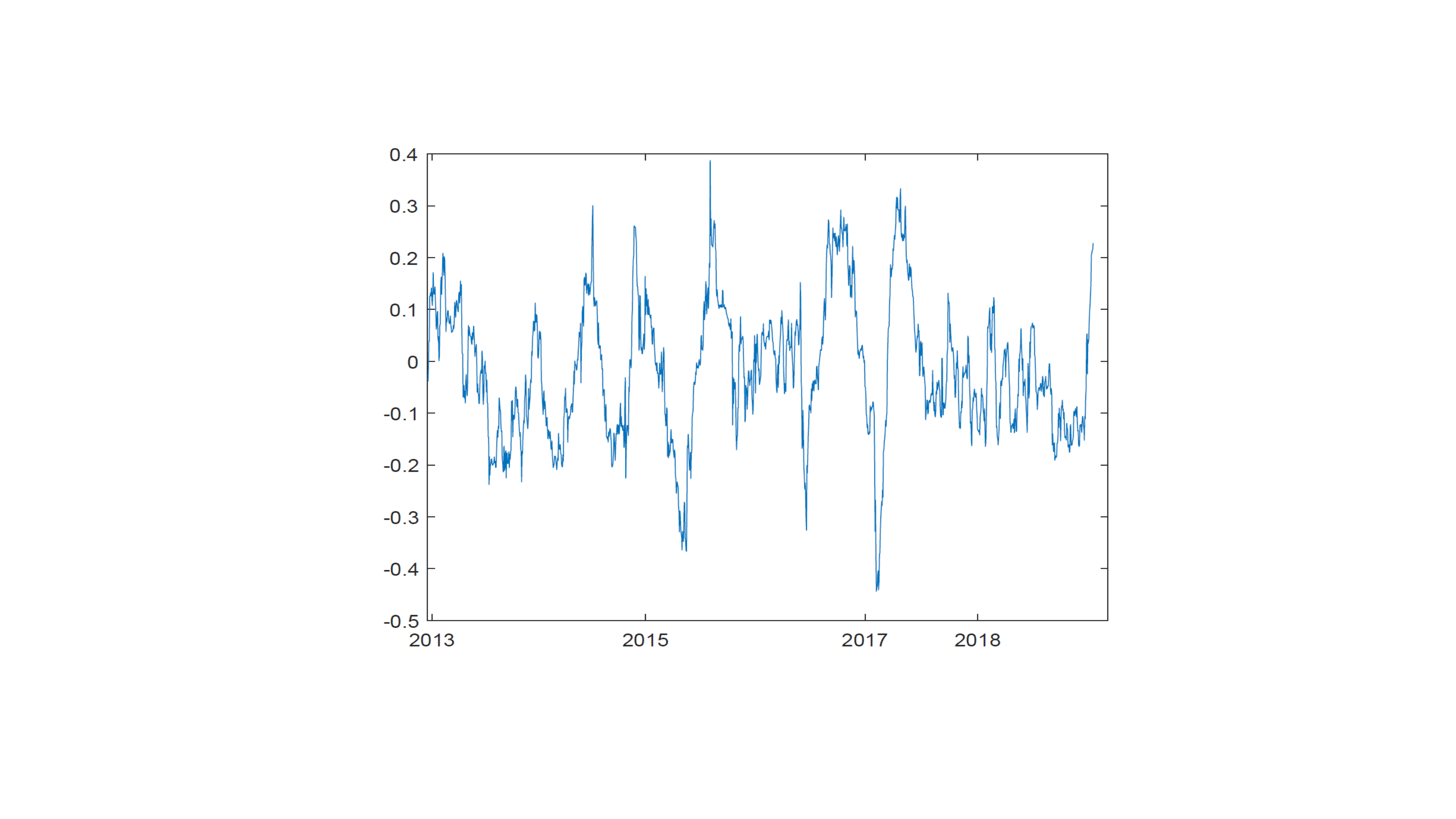}}}%\hspace{5pt}
 \caption{Left: Fifty days moving window correlation coefficient between WTI and Brent daily future prices . Right: Same window for the log-returns}\label{fig:brentwti}
\end{center}
\end{figure}
\begin{figure}[htb!]
\begin{center}
\subfigure[]{
\resizebox*{7cm}{!}{\includegraphics{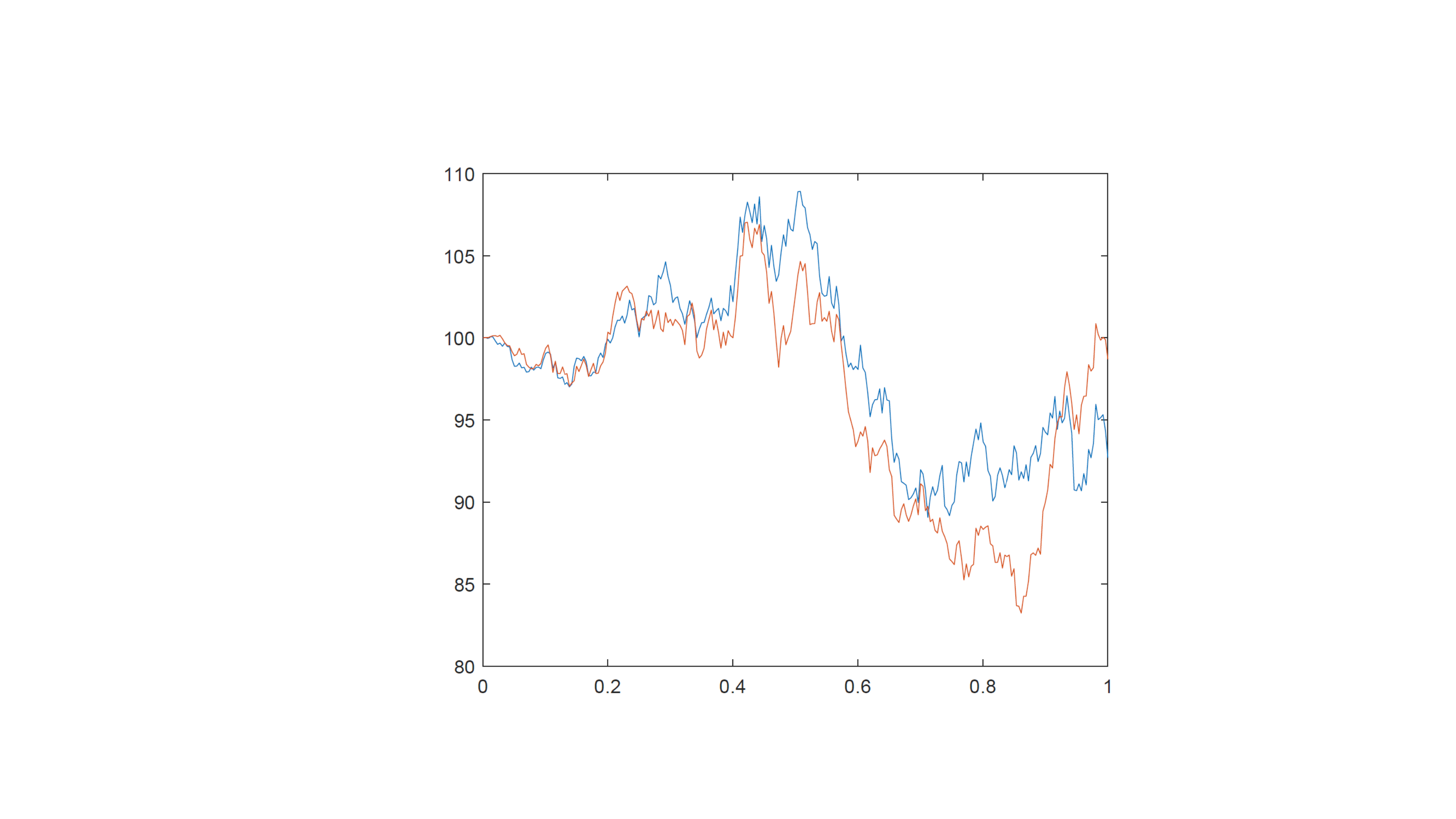}}}%\hspace{5pt}
\subfigure[]{
\resizebox*{7cm}{!}{\includegraphics{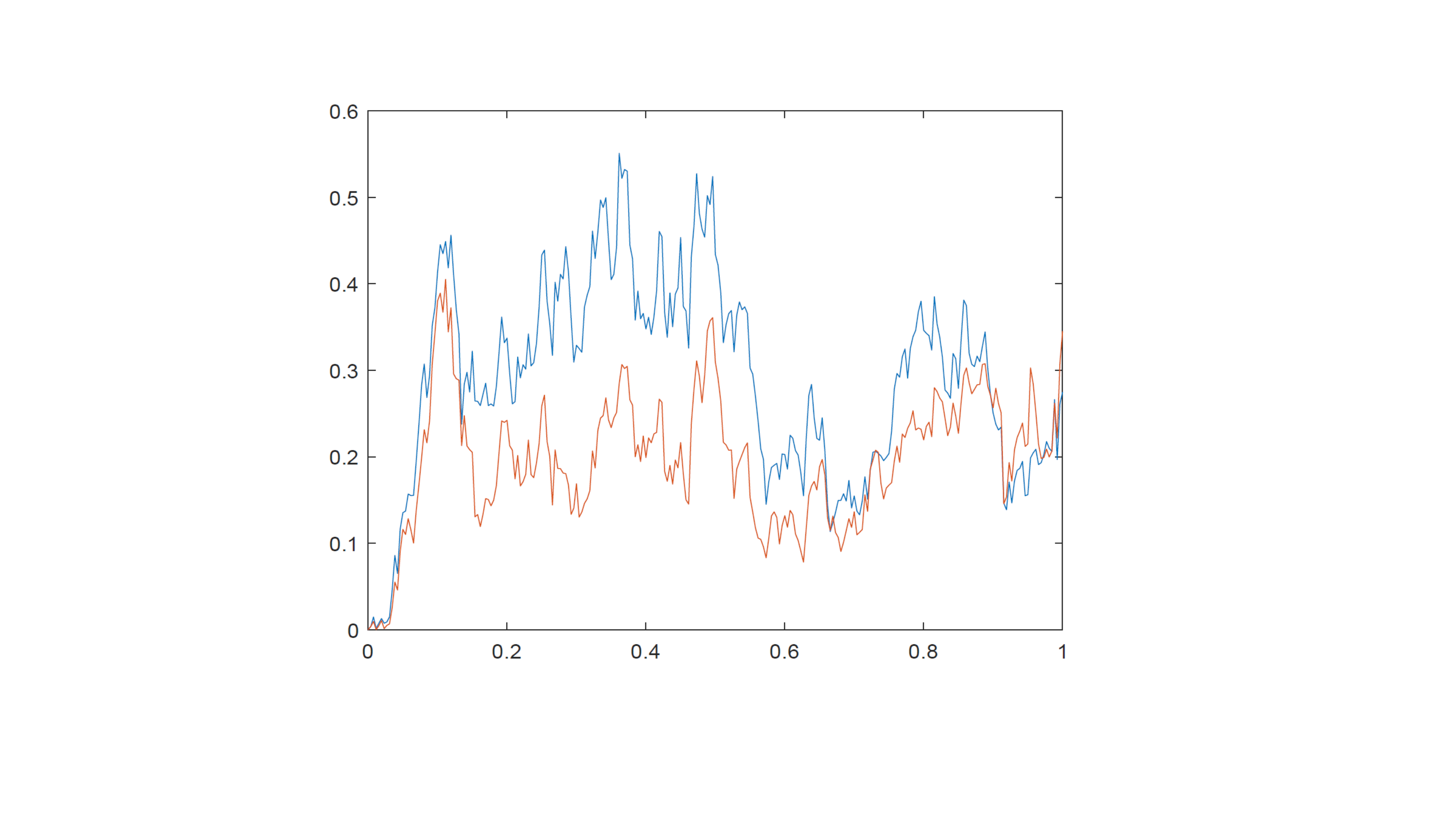}}}\hspace{5pt}
\subfigure[]{
\resizebox*{7cm}{!}{\includegraphics{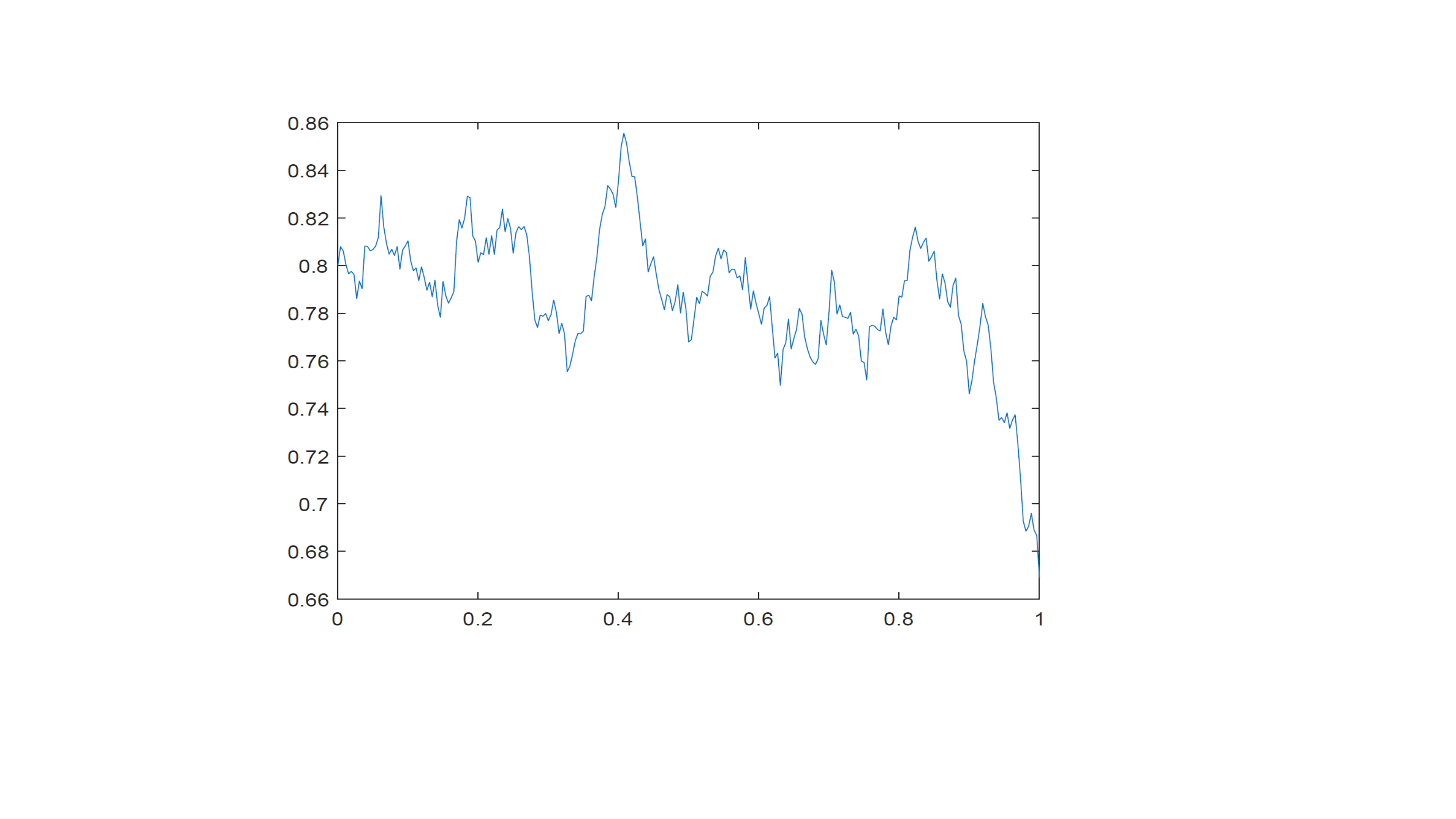}}}%\hspace{5pt}
 \caption{Counterclockwise from the top left figure a simulated  series of prices, while the top right figure shows a realization of the squared volatilities. The series in the bottom is a simulated trajectory of the correlation process.}\label{fig:simtraject}
\end{center}
\end{figure}
\begin{figure}[htb!]
\begin{center}
\subfigure[]{
\resizebox*{7cm}{!}{\includegraphics{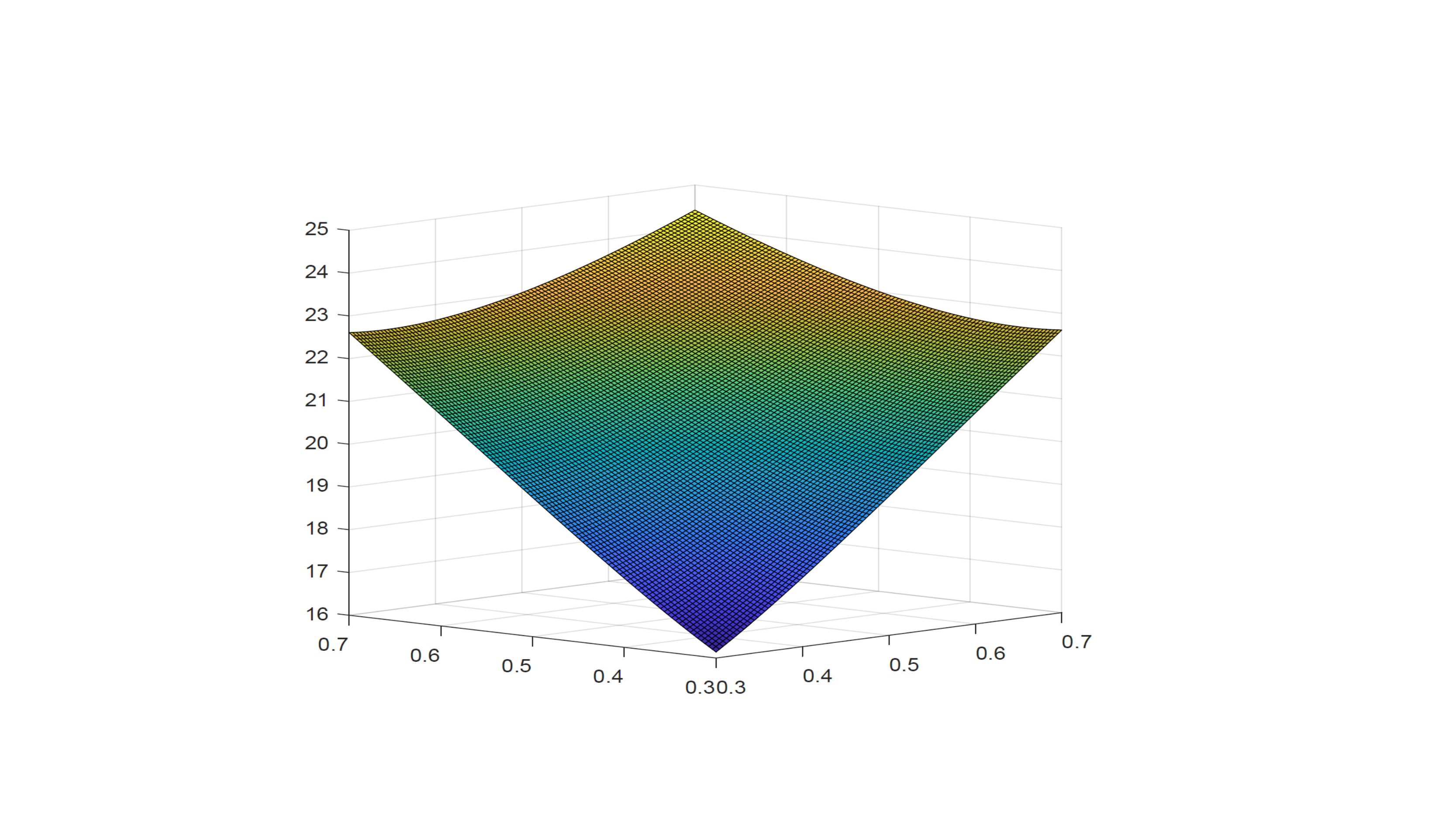}}}%\hspace{5pt}
\subfigure[]{
\resizebox*{7cm}{!}{\includegraphics{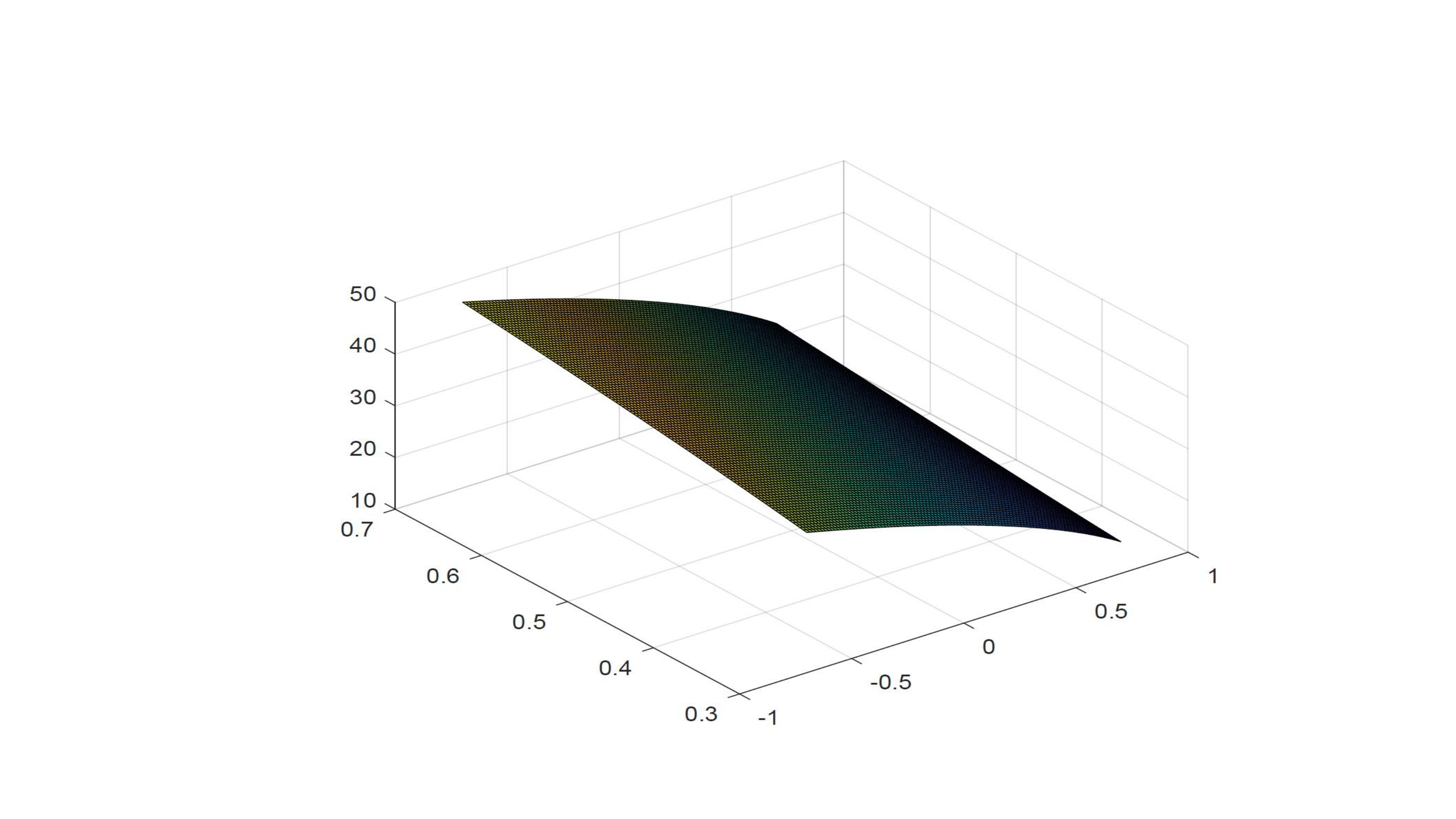}}}\hspace{5pt}

 \caption{A change in the prices of an exchange contract with respect to squared volatilities(left) and the correlation(right). }\label{fig:mcvsvol}
\end{center}
\end{figure}
\section{Numerical results}
We consider the series of  daily closure prices per barrel in US dollars in NYSE of types WTI(blue) and Brent (red), period Dec 2013 to Jan 2019 and the corresponding log-returns. Both series of prices exhibit similar patterns and, as it is expected,  are highly correlated. The overall correlation of the  series of prices is equal to $98$\% while the correlation of the log-returns is $3.81$\%. However, when the correlation is computed on a sliding windows of 50 days it exhibits notable random variations. See figures \ref{fig:brentwti}a) and b). \\
A summary of the first forth moments of the log-return series is shown in table \ref{tab:statsum}. A high kurtosis indicates the presence of  heavy-tailed distribution in both commodities.

 \begin{table}[htp!]
  \centering
  \begin{tabular}{|c|c|c|c|c|}
    \hline
  Asset&  Mean & Standard deviation & Skewness &  Kurtosis  \\  \hline
  WTI & -0.0003  &  0.0211  &  0.1089  &  6.0696   \\  \hline
  Brent   &-0.0004  &  0.0201  &  0.1473  &  5.9818 \\  \hline
   \end{tabular}
  \caption{First four moments of log-returns WTI, Brent, US/Can}\label{tab:statsum}
\end{table}
To illustrate the behavior of the components in the model we take the following set of parameters in table \ref{tab:par}. As initial prices of both assets values $S^{(1)}_0=100, S^{(2)}_0=100$ dollars are taken, initial squared volatilities $V_0=(0.3,0.3)$, the initial correlation $\rho_0=0.8$, correlation between the Brownian motions in the volatility $\rho_v=0.80$, the mean-reverting levels and rates of the volatility processes are  $V_L=(1,1)$  respectively while analogous  parameters in the correlation processes are $\bar{\Gamma}=0.8$ and $\bar{\gamma}=0.8$. The annual interest rate is $r=4$\%, and the simulation time is one year. Parameters were chosen for illustrative proposes. Other parameters are shown in table \ref{tab:par}.

 \begin{table}[htp!]
  \centering
  \begin{tabular}{|c|c|c|c|}
    \hline
  Asset& WTI  sqr. vol. & Brent sqr. vol. & Correlation \\
 Component & & &\\  \hline
   MR level  & $V^{(1)}_L=1$ & $V^{(2)}_L=1$ & $\bar{\Gamma}=0.8$ \\  \hline
    MR rate & $c_1=1$  & $c_2=1$  & $\bar{\gamma}=0.8$  \\  \hline
   vol.  & $xi_1=1$ & $xi_2=1$ & $\bar{\alpha}=1$\\  \hline
    Initial values  & $V^{(1)}_0=0.3$&  $V^{(2)}_0=0.3$ &  $\rho_0=0.7$ \\  \hline
   \end{tabular}
  \caption{Parametric set for the squared volatilities and correlation sets.}\label{tab:par}
\end{table}

The results of the simulation are shown in figure \ref{fig:simtraject}. The top left graph  represents the series of prices, while the top right figure shows a realization of the squared volatilities. The series in the bottom is a simulated trajectory of the correlation process.

A change in the prices of an exchange contract with respect to squared volatilities and the correlation are shown in figure \ref{fig:mcvsvol}. The remaining parameters are held constant. Prices are calculated according to a Monte Carlo procedure with $10^5$ realizations.

\section{Conclusions}
Taylor approximation offers a suitable method to price exchanges contracts beyond the classic framework developed originally by Margrabe. In the parametric set considered it produces accurate results with less computational effort than a traditional Monte Carlo approach.
\section{Appendix}
  \subsection{Appendix A: Moments of the volatility and correlation}
  \textbf{Proof of proposition \ref{prop:momentsrho}}
  \begin{proof}
  For the first moment notice that:
  \begin{equation}\label{eq:rholin}
     \rho_t = \rho_0+ \bar\gamma\, \bar \Gamma_L t- \bar\gamma \int_0^t \rho_s\;ds+\bar\alpha  \int_0^t \sqrt{1- \rho_s^2 }\,d\bar W_s
  \end{equation}
  Taking expected value on both sides:
  \begin{equation*}
   mr_1(t):= E_{\mathcal{Q}}(\rho_t) = \rho_0+ \bar\gamma\,\bar \Gamma_L t- \bar\gamma \int_0^t  mr_1(s) \;ds
  \end{equation*}
  Differentiating we get:
   \begin{equation*}
   mr_1'(t)=  \bar\gamma \bar \Gamma_L -  \bar\gamma  mr_1(t)
  \end{equation*}
  whose solution is:
  \begin{equation*}
   mr_1(t)=   \bar \Gamma_L + (\rho_0-\bar \Gamma_L)e^{-\bar\gamma t}
  \end{equation*}
  Similarly, for the integrated process:
  \begin{eqnarray*}
  E_{\mathcal{Q}}(\rho^+_t) &=&  \int_0^t \bar \Gamma_L + (\rho_0-\bar \Gamma_L)e^{-\bar\gamma s}\;ds\\
  &=& \bar \Gamma_L t+\left(\frac{\rho_0-\bar \Gamma_L}{\bar\gamma}\right)(1-e^{-\bar\gamma t})
  \end{eqnarray*}
  To compute the second moment we first apply Ito formula to $f(x)=x^2$ and the correlation process. Hence:
  \begin{eqnarray*}
  \rho^2_t &=& \rho^2_0+ 2 \int_0^t \rho_s d \rho_s+ <\rho_t>\\
   &=& \rho^2_0+ 2 \bar\gamma \bar \Gamma_L \int_0^t \rho_s\;ds - 2 \bar\gamma \int_0^t \rho^2_s\;ds+2 \bar\alpha  \int_0^t \rho_s \sqrt{1- \rho_s^2 }\,d\bar W_s\\
   &+& \bar\alpha^2 \int_0^t  (1- \rho_s^2 )\,ds\\
  &=&  \rho^2_0+ \bar\alpha^2 t + 2 \bar\gamma \bar \Gamma_L \int_0^t \rho_s\;ds-(\bar\alpha^2+2 \bar\gamma)\int_0^t \rho^2_s\;ds \\
  &+& 2 \bar\alpha  \int_0^t \rho_s \sqrt{1- \rho_s^2 }\,d\bar W_s
    \end{eqnarray*}
  Taking expected value:
  \begin{eqnarray*}
  E_{\mathcal{Q}}(\rho^2_t) &=&  \rho^2_0+ \bar\alpha^2 t + 2 \bar\gamma \bar \Gamma_L \int_0^t E_{\mathcal{Q}}(\rho_s)\;ds-(\bar\alpha^2+2 \bar\gamma)\int_0^t E_{\mathcal{Q}}(\rho^2_s)\;ds
   \end{eqnarray*}
   or after differentiating:
   \begin{eqnarray*}
     mr'_2(t)&+& (2\bar\gamma+\bar\alpha^2 )mr_2(t) =  2 \bar\gamma \bar \Gamma_L mr_1(t)+   \bar\alpha^2 \\
  mr_2(0) &=& \rho^2_0
     \end{eqnarray*}
   its solution is:
      \begin{eqnarray}\label{eq:smomrho}
  mr_2(t) &=& a_1 + a_2 e^{-\bar\gamma t}+ (\rho^2_0-a_1-a_2)  e^{-(2\bar\gamma+\bar\alpha^2)t}
   \end{eqnarray}
    Next, notice that we have:
  \begin{eqnarray*}
  \frac{mr^{+}_{2}}{dt} &=& 2 E_{\mathcal{Q}}[\rho^+_t \rho_t]
  \end{eqnarray*}
  From  equation (\ref{eq:rholin}):
  \begin{eqnarray*}
   E_{\mathcal{Q}}(\rho_t+\bar\gamma \rho^+_t)^2&=&  E_{\mathcal{Q}}(\rho_0+ \bar\gamma\, \bar \Gamma_L t+\bar\alpha  \int_0^t \sqrt{1- \rho_s^2 }\,d\bar W_s)^2
  \end{eqnarray*}
  Expanding both  sides in the equation above we have:
   \begin{eqnarray*}
   LHS=E_{\mathcal{Q}}(\rho_t+\bar\gamma \rho^+_t)^2&=& E_{\mathcal{Q}}(\rho^2_t)+2 \bar\gamma E_{\mathcal{Q}}(\rho_t \rho^+_t)+\bar\gamma^2 E_{\mathcal{Q}}( \rho^+_t)^2\\
   &=& mr_2(t)+ \bar\gamma \frac{mr^{+}_{2}}{dt}+\bar\gamma^2 mr^+_{2}(t)
  \end{eqnarray*}
  and
   \begin{eqnarray*}
   RHS&=&  (\rho_0+ \bar\gamma\, \bar \Gamma_L t)^2+2 (\rho_0+ 2 \bar\gamma\, \bar \Gamma_L t)\bar\alpha  E_{\mathcal{Q}}(\int_0^t \sqrt{1- \rho_s^2 }\,d\bar W_s)\\
   &+& \bar\alpha^2   E_{\mathcal{Q}} \left(\int_0^t \sqrt{1- \rho_s^2 }\,d\bar W_s \right)^2\\
   &=& (\rho_0+ \bar\gamma\, \bar \Gamma_L t)^2+  \bar\alpha^2   E_{\mathcal{Q}}(\int_0^t \sqrt{1- \rho_s^2 }\,d\bar W_s)^2\\
   &=& (\rho_0+ \bar\gamma\, \bar \Gamma_L t)^2+  \bar\alpha^2   E_{\mathcal{Q}} \left(\int_0^t (1- \rho_s^2)\,ds \right)\\
   &=& (\rho_0+ \bar\gamma\, \bar \Gamma_L t)^2+  \bar\alpha^2  (t- \int_0^t mr_2(s)\;ds)
  \end{eqnarray*}
  From equation (\ref{eq:smomrho}):
  \begin{eqnarray*}
 \int_0^t mr_2(s)\;ds &=&  \int_0^t (a_1 + a_2 e^{-\bar\gamma s}+ (\rho^2_0-a_1-a_2)  e^{-(2\bar\gamma+\bar\alpha^2)s}\;ds\\
 &=& a_1 t + \frac{a_2}{\bar\gamma}(1-e^{-\bar\gamma t})+ \frac{\rho^2_0-a_1-a_2}{2\bar\gamma+\bar\alpha^2}(1-e^{-(2\bar\gamma+\bar\alpha^2)t})
   \end{eqnarray*}
  Hence,
  \begin{eqnarray*}
 \frac{mr^{+}_{2}}{dt} + \bar\gamma mr^+_{2}(t)&=& b(t)
  \end{eqnarray*}
  where:
   \begin{eqnarray*}
b(t)&=&  \frac{1}{\bar\gamma}\left(-mr_2(t)+(\rho_0+ \bar\gamma\, \bar \Gamma_L t)^2+ \bar\alpha^2 t-\frac{\bar\alpha^2}{\bar\gamma}\int_0^t mr_2(s)\;ds \right)\\
&=& \frac{1}{\bar\gamma}\left(-mr_2(t)+(\rho_0+ \bar\gamma\, \bar \Gamma_L t)^2+ \bar\alpha^2 t \right. \\
&-& \frac{1}{\bar\gamma} \left. \left(a_1 t + \frac{a_2}{\bar\gamma}(1-e^{-\bar\gamma t})+ \frac{\rho^2_0-a_1-a_2}{2\bar\gamma+\bar\alpha^2}(1-e^{-(2\bar\gamma+\bar\alpha^2)t}) \right) \right)
  \end{eqnarray*}
  and initial condition $mr_{+,2}(0)=0$.\\
  Using the integrating factor $e^{\bar\gamma t}$ we find that its solution is:
  \begin{eqnarray}\label{eq:bt}
  mr^+_{2}(t) &=&  e^{-\bar\gamma t}\int e^{\bar\gamma t}b(t)\;dt+c e^{-\bar\gamma t}
  \end{eqnarray}
  But:
  \begin{eqnarray*}
 \int e^{\bar\gamma t}b(t)\;dt  &=& \frac{1}{\bar\gamma} \int e^{\bar\gamma t} \left(-mr_2(t)+(\rho_0+ \bar\gamma\, \bar \Gamma_L t)^2+ \bar\alpha^2 t \right) \\
&-& \frac{\bar\alpha^2}{\bar\gamma^2} \int e^{\bar\gamma t}  \left(a_1 t + \frac{a_2}{\bar\gamma}(1-e^{-\bar\gamma t})+ \frac{\rho^2_0-a_1-a_2}{2\bar\gamma+\bar\alpha^2}(1-e^{-(2\bar\gamma+\bar\alpha^2)t}) \right) \;dt
  \end{eqnarray*}
Moreover, from equation (\ref{eq:smomrho}):
\begin{eqnarray*}
 \int e^{\bar\gamma t} mr_2(t)\;dt &=&  \int e^{\bar\gamma t} (a_1 + a_2 e^{-\bar\gamma t}+ (\rho^2_0-a_1-a_2)  e^{-(2\bar\gamma+\bar\alpha^2)t}) \;dt\\
 &=& \frac{a_1}{\bar\gamma}e^{\bar\gamma t}  + a_2 t - \frac{\rho^2_0-a_1-a_2}{\bar\gamma+\bar\alpha^2}  exp(-(\bar\gamma+\bar\alpha^2)t)) \\
  \int (\rho_0+ \bar\gamma\, \bar \Gamma_L t)^2e^{\bar\gamma t}\;dt &=& \frac{\rho^2_0}{\bar\gamma}e^{\bar\gamma t}+2 \bar\gamma \bar \Gamma_L (\frac{1}{\bar\gamma}e^{\bar\gamma t}t-\frac{1}{\bar\gamma^2}e^{\bar\gamma t})\\
 &+& \bar\gamma^2 \bar \Gamma^2_L(\frac{1}{\bar\gamma}t^2e^{\bar\gamma t}-\frac{2}{\bar\gamma}t e^{\bar\gamma t}+\frac{2}{\bar\gamma^3} e^{\bar\gamma t})\\
 &=& [\rho^2_0+2 \bar\gamma \bar \Gamma_L(t-\frac{1}{\bar\gamma})+t^2-\frac{2}{\bar\gamma}t +\frac{2}{\bar\gamma^2}] \frac{1}{\bar\gamma}e^{\bar\gamma t}\\
 \int t e^{\bar\gamma t} \;dt &=& [t-\frac{1}{\bar\gamma^2}] \frac{1}{\bar\gamma}e^{\bar\gamma t}
\end{eqnarray*}
Hence:
\begin{eqnarray*}
\int e^{\bar\gamma t}b(t)\;dt &=& -\frac{1}{\bar\gamma^2}a_1 e^{\bar\gamma t}-(\frac{a_2}{\bar\gamma})t + \left(\frac{\rho^2_0-a_1-a_2}{\bar\gamma(\bar\gamma+\bar\alpha^2)} \right)  e^{-(\bar\gamma+\bar\alpha^2)t}\\
&+& \frac{1}{\bar\gamma^2}(\rho^2_0+2 \bar\gamma \bar \Gamma_L(t-\frac{1}{\bar\gamma})+t^2-\frac{2}{\bar\gamma}t +\frac{2}{\bar\gamma^2}) e^{\bar\gamma t}\\
&+& \frac{\bar\alpha^2}{\bar\gamma^2}\left(t- \frac{1}{\bar\gamma^2} \right) e^{\bar\gamma t}\\
 &-& \frac{\bar\alpha^2}{\bar\gamma^2} \left( \frac{a_1}{\bar\gamma}(t-\frac{1}{\bar\gamma^2})e^{\bar\gamma t}+ \frac{a_2}{\bar\gamma^2}e^{\bar\gamma t}- \frac{a_2}{ \bar\gamma}t \right.\\
 &+& \left. \frac{\rho^2_0-a_1-a_2}{ \bar\gamma(2 \bar\gamma+\bar\alpha^2)}e^{\bar\gamma t}+\frac{\rho^2_0-a_1-a_2}{(2 \bar\gamma+\bar\alpha^2)(\bar\gamma+\bar\alpha^2)}e^{-(\bar\gamma+\bar\alpha^2)t}\right)
\end{eqnarray*}
Combining the expressions above into equation (\ref{eq:bt}) we have:
\begin{eqnarray*}
  mr^+_{2}(t) &=&  b_0+b_1t+b_2 t^2+b_3 t e^{-\bar\gamma t}+b_4 e^{-(2\bar\gamma+\bar\alpha^2)t}+c e^{-\bar\gamma t}
   \end{eqnarray*}
  From the initial conditions $c=-b_0-b_4$.\\
    Combining the first and second moments of $\rho^+_t$ we obtain the expression for the variance in equation \ref{eq:varrho}.
  \end{proof}
    \textbf{Proof of proposition \ref{prop:momentsv}}
  \begin{proof}
  To compute the first and second moments we proceed similarly to the proof of proposition 2.3. Notice equations for squared volatilities are of mean-reverting square root type s.d.e's as well.\\
 Hence:
\begin{eqnarray*}
mv_{1,j}(t) &=& V^{(j)}_L +(V^{(j)}_0-V^{(j)}_L)e^{-c_jt}\\
mv^+_{1,j}(t)&=& E_{\mathcal{Q}}[V_t^{+,j}]= V^{(j)}_L t +\frac{V^{(j)}_0-V^{(j)}_L}{c_j}(1-e^{-c_jt})
\end{eqnarray*}
Moreover,
\begin{eqnarray*}
  (V^{(j)}_t)^2 &=& (V^{(j)}_0)^2+ 2 \int_0^t V^{(j)}_s d V^{(j)}_s+ <V^{(j)}_t>\\
   &=& (V^{(j)}_0)^2+ 2 c_j V^{(j)}_L  V^{j,+}_t - 2 c_j \int_0^t (V^{(j)}_s)^2\;ds+2 \xi_j  \int_0^t V^{(j)}_s \sigma^{(j)}_s \,dW^{(j)}_s\\
   &+& \xi_j^2 V^{j,+}_t\\
  &=&  (V^{(j)}_0)^2+  (2 c_j V^{(j)}_L+\xi^2_j) V^{j,+}_t-2 c_j \int_0^t (V^{(j)}_s)^2\;ds \\
  &+& 2 \xi_j  \int_0^t V^{(j)}_s \sigma^{(j)}_s\,dW^{(j)}_s
    \end{eqnarray*}
  Taking expected value on both sides:
  \begin{eqnarray*}
 mv_{2,j}(t) &=&   (V^{(j)}_0)^2+  (2 c_j V^{(j)}_L+\xi^2_j) \int_0^t mv_{1,j}(s)\;ds-2 c_j \int_0^t mv_{2,j}(s)\;ds
   \end{eqnarray*}
   or
   \begin{eqnarray*}
  mv'_{2,j}(t)&+& 2c_j mv_{2,j}(t)=(2 c_j V^{(j)}_L+\xi^2_j)mv_{1,j}(t)   \\
  mv_{2,j}(0) &=& (V^{(j)}_0)^2
   \end{eqnarray*}
   with $c(t)=(2 c_j V^{(j)}_L+\xi_j^2)mv_{1,j}(t)$.\\
      Its solution is:
    \begin{eqnarray*}
   mv_{2,j}(t) &=&   e^{-2c_j t} \int  e^{2 c_j t}c(t)\;dt+d_2 e^{-2c_j t}
    \end{eqnarray*}
   But:
    \begin{eqnarray*}
   \int  e^{2 c_j t}c(t)\;dt &=& (2 c_j V^{(j)}_L+\xi_j^2) \int  e^{2 c_j t}mv_{1,j}(t) \;dt\\
   &=&  (2 c_j V^{(j)}_L+\xi_j^2) \int  e^{2 c_j t}(V^{(j)}_L +(V^{(j)}_0-V^{(j)}_L)e^{-c_jt}) \;dt\\
      &=&  (2 c_j V^{(j)}_L+\xi_j^2) (\frac{V^{(j)}_L}{2 c_j}e^{2 c_j t}+\frac{V^{(j)}_0-V^{(j)}_L}{c_j}e^{c_j t})
  \end{eqnarray*}
  Then:
      \begin{eqnarray}\label{eq:smomrho}
  mv_{2,j}(t) &=& d_0 + d_1 e^{-c_j t}+ d_2  e^{-2c_j t}
   \end{eqnarray}
   where:
   \begin{eqnarray*}
    d_0 &=& (2 c_j V^{(j)}_L+\xi_j^2)\frac{ V^{(j)}_L}{2 c_j}\\
   d_1 &=& (2 c_j + \xi_j^2)\frac{(V^{(j)}_0-V^{(j)}_L)}{c_j}\\
   d_2 &=& (V^{(j)}_0)^2-d_0-d_1
   \end{eqnarray*}
  Next, notice that we have:
  \begin{eqnarray}\label{eq:rhorhomas}
  \frac{dmv^+_{2,j}}{dt} &=& 2 E_{\mathcal{Q}}[V^{(j,+)}_t V^{(j)}_t]
  \end{eqnarray}
  Now:
  \begin{eqnarray}\nonumber
   E_{\mathcal{Q}}(V^{(j)}_t+c_j V^{j,+}_t)^2&=&  E_{\mathcal{Q}}[V^{(j)}_0+ c_j\, V^{(j)}_L t+\xi_j  \int_0^t\sigma^{(j)}_s \,dW^{(j)}_s]^2\\ \nonumber
   &=& (V^{(j)}_0+ c_j\, V^{(j)}_L t)^2+2(V^{(j)}_0+c_j  V^{(j)}_L t)  \xi_j  E_{\mathcal{Q}}\left(\int_0^t \sigma^{(j)}_s \,dW^{(j)}_s \right)\\ \nonumber
   &+& \xi^2_j  E_{\mathcal{Q}}\left(\int_0^t \sigma^{(j)}_s \,dW^{(j)}_s \right)^2\\ \label{eq:rhs}
    &=& (V^{(j)}_0+ c_j\, V^{(j)}_L t)^2+ \xi^2_j  \int_0^t mv_{1,j}(s)\;ds
  \end{eqnarray}
On the other hand, after expanding the expression above and taking into account equation (\ref{eq:rhorhomas}):
 \begin{eqnarray}\label{eq:mv2plus}
   E_{\mathcal{Q}}(V^{(j)}_t+c_j V^{j,+}_t)^2&=& mv_{2,j}(t)+c_j\frac{dmv^+_{2,j}}{dt}+ c^2_j mv^+_{2,j}
    \end{eqnarray}
  Hence, equating equations (\ref{eq:rhs}) and (\ref{eq:rhorhomas}) we have that  $mv^+_{2,j}$ satisfies:
 \begin{eqnarray} \label{eq:intvsqvol}
 \frac{d mv^+_{2,j}}{dt}+ c_j mv^+_{2,j}(t) &=&d(t)\\ \nonumber
 mv^+_{2,j}(0) &=& 0
  \end{eqnarray}
  with:
  \begin{eqnarray*}
  d(t)&=& \frac{(V^{(j)}_0+ c_j\, V^{(j)}_L t)^2}{c_j}+ \frac{\xi^2_j}{c_j}  \int_0^t mv_{1,j}(s)\;ds -\frac{1}{c_j} mv_{2,j}(t)\\
  &=&\frac{(V^{(j)}_0+ c_j\, V^{(j)}_L t)^2}{c_j}+  \frac{\xi^2_j}{c_j} \int_0^t  V^{(j)}_L +(V^{(j)}_0-V^{(j)}_L)e^{-c_jt}\;ds \\
  &-& \frac{1}{c_j}(d_0 + d_1 e^{-c_j t}+ d_2  e^{-2c_j t})\\
  &=&\frac{(V^{(j)}_0+ c_j\, V^{(j)}_L t)^2}{c_j}+   \frac{\xi^2_j}{c_j}  V^{(j)}_L t -\frac{\xi^2_j }{c^2_j}(V^{(j)}_0-V^{(j)}_L)e^{-c_jt} \\
  &-& \frac{1}{c_j}(d_0 + d_1 e^{-c_j t}+ d_2  e^{-2c_j t})
    \end{eqnarray*}
  The solution of equation (\ref{eq:intvsqvol}) is:
  \begin{eqnarray}\nonumber
    mv^+_{2,j}(t) &=&  e^{-c_j t} \int e^{c_j t}d(t)\;dt+c e^{-c_j t}\\ \nonumber
    &=& e^{-c_j t} \int e^{c_j t}\left[\frac{(V^{(j)}_0+ c_j\, V^{(j)}_L t)^2}{c_j}+   \frac{\xi^2_j}{c_j}  V^{(j)}_L t -\frac{\xi^2_j }{c^2_j}(V^{(j)}_0-V^{(j)}_L)e^{-c_jt} \right. \\ \nonumber
  &-& \left. \frac{1}{c_j}(d_0 + d_1 e^{-c_j t}+ d_2  e^{-2c_j t}) \right]\;dt+c e^{-c_j t}\\ \nonumber
   &=& \frac{e^{-c_j t}}{c_j} \int e^{c_j t}\left[(V^{(j)}_0+ c_j\, V^{(j)}_L t)^2+   \xi^2_j  V^{(j)}_L t -\frac{\xi^2_j }{c_j}(V^{(j)}_0-V^{(j)}_L)e^{-c_jt} \right. \\ \nonumber
  &-& \left. (d_0 + d_1 e^{-c_j t}+ d_2  e^{-2c_j t} ) \right]\;dt+d_3 e^{-c_j t}\\ \label{eq:diff2mom}
  &&
    \end{eqnarray}
    Moreover:
    \begin{eqnarray*}
    && \int e^{c_j t}(V^{(j)}_0+ c_j\, V^{(j)}_L t)^2\;dt = \frac{(V^{(j)}_0)^2}{c_j}V^{(j)}_0 e^{c_j t}+ 2 c_j\, V^{(j)}_L \int t e^{c_j t} \;dt + (V^{(j)}_L )^2 c^2_j \int t^2 e^{c_j t} \;dt\\
    &=& \frac{(V^{(j)}_0)^2}{c_j}e^{c_j t}+ 2 c_j V^{(j)}_0 V^{(j)}_L (\frac{t}{c_j}e^{c_j t}-\frac{1}{c^2_j}e^{c_j t})+ c^2_j (V^{(j)}_L )^2 (\frac{t^2}{c_j}e^{c_j t}- \frac{2 t}{c^2_j}e^{c_j t}+\frac{2 }{c^3_j}e^{c_j t})\\
    &=& \frac{e^{c_j t}}{c_j}\left[c^2_j(V^{(j)}_L)^2 t^2+2( c_j V^{(j)}_0 V^{(j)}_L- c_j (V^{(j)}_L)^2) t + (V^{(j)}_0)^2+2(V^{(j)}_L)^2 \right]
     \end{eqnarray*}

    \begin{eqnarray*}
     \int e^{c_j t} \xi^2_j  V^{(j)}_L t \;dt  &=& \frac{\xi^2_j  V^{(j)}_L}{c_j}e^{c_j t}(t-\frac{1}{c_j})\\
     \int e^{c_j t} \frac{\xi^2_j }{c_j}(V^{(j)}_0-V^{(j)}_L)e^{-c_jt} \;dt  &=& \frac{\xi^2_j }{c_j}(V^{(j)}_0-V^{(j)}_L)t\\
     \int e^{c_j t}  (d_0 + d_1 e^{-c_j t}+ d_2  e^{-2c_j t}) \;dt  &=& \frac{d_0}{c_j} e^{c_j t}\\
    &+& d_1 t-\frac{d_2}{c_j}e^{-c_j t}\\
    &=&  -\frac{e^{-c_j t}}{c_j}[d_0+(V^{(j)}_0)^2-d_1+\frac{d_1}{2}e^{-c_j t}]
       \end{eqnarray*}
Therefore substituting in equation (\ref{eq:diff2mom}):
 \begin{eqnarray*}
    mv^+_{2,j}(t) &=&  \frac{1}{c^3_j}\left[(V^{(j)}_L)^2 t^2+(2 c_j - \frac{V^{(j)}_L}{c_j})V^{(j)}_L t + (V^{(j)}_0)^2+\frac{2(V^{(j)}_L)^2}{c^2_j} \right]\\
  &+& \frac{\xi^2_j  V^{(j)}_L}{c^3_j}(t-\frac{1}{c_j})+ \frac{\xi^2_j e^{-c_j t}}{c^3_j}(V^{(j)}_0-V^{(j)}_L)t\\
  &-& \frac{e^{-2c_j t}}{c^3_j}[d_0+(V^{(j)}_0)^2-d_1+\frac{d_1}{2}e^{-c_j t}]   +d_3 e^{-c_j t}
    \end{eqnarray*}
    Where, from the initial conditions:
    \begin{equation*}
      d_3=\frac{1}{c^3_j}[\frac{1}{c_j}\xi^2_j  V^{(j)}_L+d_0+\frac{d_1}{2}-\frac{2(V^{(j)}_L)^2 }{c^2_j}]
    \end{equation*}
    \end{proof}
    \textbf{Proof of proposition \ref{eq:covvol}}
    \begin{proof}
To compute the covariance of the integrated squared volatilities we start noticing that $ <\sigma_t^{(1)}, \sigma_t^{(2)}>=\beta_1 \beta_2 \rho_V t$.
Therefore by integration by parts formula:
  \begin{eqnarray*}
 \sigma_t^{(1)} \sigma_t^{(2)} &=& \sigma_0^{(1)} \sigma_0^{(2)}+ \int_0^t \sigma_s^{(1)} d \sigma_s^{(2)}+ \int_0^t \sigma_s^{(2)} d \sigma_t^{(1)}+ <\sigma_t^{(1)}, \sigma_t^{(2)}>\\
 &=& \sigma_0^{(1)} \sigma_0^{(2)}- \alpha_2 \int_0^t \sigma_s^{(1)}  \sigma_s^{(2)} ds+ \beta_2 \int_0^t \sigma_s^{(1)}dW_t^{(2)}  \\
 &-& \alpha_1 \int_0^t \sigma_s^{(1)}  \sigma_s^{(2)} ds+ \beta_1 \int_0^t \sigma_s^{(2)}dW_t^{(1)}+ \beta_1 \beta_2 \rho_V t
   \end{eqnarray*}
   Taking expected value on both sides:
   \begin{eqnarray*}
 E_{\mathcal{Q}}[\sigma_t^{(1)} \sigma_t^{(2)}] &=& \sigma_0^{(1)} \sigma_0^{(2)}- (\alpha_1+\alpha_2) \int_0^t E_{\mathcal{Q}}[\sigma_s^{(1)}  \sigma_s^{(2)}]\;ds +  \beta_1 \beta_2 \rho_V t
   \end{eqnarray*}
   The expression above leads to the differential equation:
   \begin{equation*}
     ms'_{12}(t)+ (\alpha_1+\alpha_2)ms_{12}(t)-\beta_1 \beta_2 \rho_V=0
   \end{equation*}
   with $ms_{12}(t)= E_{\mathcal{Q}}[\sigma_t^{(1)} \sigma_t^{(2)}]$.\\
    Its solution is:
   \begin{equation*}
     ms_{12}(t)=\frac{\beta_1 \beta_2 \rho_V}{\alpha_1+\alpha_2}\left( 1-e^{-(\alpha_1+\alpha_2)t} \right)+\sigma_0^{(1)} \sigma_0^{(2)}e^{-(\alpha_1+\alpha_2)t}
   \end{equation*}
   With the reparametrization in remark \ref{rem:par} it becomes:
    \begin{equation}\label{eq:covsigmas}
     ms_{12}(t)=\frac{\xi_1 \xi_2 \rho_V}{2(c_1+c_2}\left( 1-\exp(-\frac{1}{2}(c_1+c_2)t) \right)+\sigma_0^{(1)} \sigma_0^{(2)}\exp(-\frac{1}{2}(c_1+c_2)t)
   \end{equation}
   Moreover, from equation (\ref{eq: svol1}):
    \begin{eqnarray*}
 V_t^{(1)} V_t^{(2)} &=& V_0^{(1)} V_0^{(2)}+ \int_0^t V_s^{(1)} d V_s^{(2)}+ \int_0^t V_s^{(2)} d V_s^{(1)}+ <V_t^{(1)}, V_t^{(2)}>\\
 &=& V_0^{(1)} V_0^{(2)}+c_2 V^{(2)}_L t-c_2 \int_0^t V_s^{(1)}  V_s^{(2)} ds+ \xi_2 \int_0^t V_s^{(1)}\sigma_s^{(2)}dW_t^{(2)}  \\
 &+& c_1 V^{(1)}_L t-c_1 \int_0^t V_s^{(2)}  V_s^{(1)} ds+ \xi_1 \int_0^t V_s^{(2)}\sigma_s^{(1)}dW_t^{(1)}+ \xi_1 \xi_2 \rho_V \int_0^t \sigma_s^{(1)}\sigma_s^{(2)}ds
   \end{eqnarray*}
   Again, taking expected value on both sides of the equation above and differentiating:
   \begin{equation*}
     mv'_{12}(t)=c_1V^{(1)}_L +c_2 V^{(2)}_L-(c_1+c_2)mv_{12}(t)+\xi_1 \xi_2 \rho_V m_{12}(t)
   \end{equation*}
   whose solution is given by:
   \begin{eqnarray*}
   mv_{12}(t) &=&  e^{-(c_1+c_2)t}\xi_1 \xi_2 \rho_V \int e^{(c_1+c_2)s}ms_{12}(s)\;ds+ c e^{-(c_1+c_2)t}\\
   &=& e^{-(c_1+c_2)t}\xi_1 \xi_2 \rho_V \int e^{(c_1+c_2)s}[\frac{\xi_1 \xi_2 \rho_V}{2(c_1+c_2}\left( 1-e^{-\frac{1}{2}(c_1+c_2)s} \right)\;ds\\
   &+&\sigma_0^{(1)} \sigma_0^{(2)}\xi_1 \xi_2 \rho_V e^{-(c_1+c_2)t} \int e^{(c_1+c_2)s} e^{-\frac{1}{2}(c_1+c_2)s}\;ds + c e^{-(c_1+c_2)t}\\
   &=& \frac{(\xi_1 \xi_2 \rho_V)^2}{2(c_1+c_2)}e^{-(c_1+c_2)t} \int e^{(c_1+c_2)s}\;ds- \frac{(\xi_1 \xi_2 \rho_V)^2}{2(c_1+c_2)}e^{-(c_1+c_2)t} \int \exp(\frac{1}{2}(c_1+c_2)s)\;ds\\
   &+&\sigma_0^{(1)} \sigma_0^{(2)}\xi_1 \xi_2 \rho_V e^{-(c_1+c_2)t} \int e^{\frac{1}{2}(c_1+c_2)s}\;ds + c e^{-(c_1+c_2)t}\\
   &=& \frac{(\xi_1 \xi_2 \rho_V)^2}{2(c_1+c_2)^2}- \frac{(\xi_1 \xi_2 \rho_V)^2}{(c_1+c_2)^2}e^{-\frac{1}{2}(c_1+c_2)t}+ \frac{2\sigma_0^{(1)} \sigma_0^{(2)}\xi_1 \xi_2 \rho_V}{c_1+c_2} e^{-\frac{1}{2}(c_1+c_2)t}  + c e^{-(c_1+c_2)t}
   \end{eqnarray*}
   From the initial condition $ mv_{12}(0)=V_0^{(1)} V_0^{(2)}$ we have that:
   \begin{equation*}
     c=V_0^{(1)} V_0^{(2)}+\frac{1}{2}\frac{(\xi_1 \xi_2 \rho_V)^2}{(c_1+c_2)^2}-\frac{2\sigma_0^{(1)} \sigma_0^{(2)}\xi_1 \xi_2 \rho_V}{c_1+c_2}
   \end{equation*}
   On the other hand, from equation (\ref{eq: svol1}):
   \begin{eqnarray*}
   V^{j,+}_t&=& \frac{1}{c_j}[V^{(j)}_0+c_1 V^{(j)}_L t-V^{(j)}_t +\xi_j  \sigma^{(j)}_t\, dW_t^{(j)}]\\
  V^{(j)}_t &=& V^{(j)}_0 e^{-c_j t}+V^{(j)}_L(1-e^{-c_j t})+\xi_j e^{-c_j t} \int_0^t e^{c_j s}\sigma^{(j)}_s \, dW_t^{(j)}
   \end{eqnarray*}
   Hence:
   \begin{eqnarray*}
   mv^+_{12}(t) &:=& E_{\mathcal{Q}}[V^{1,+}_t V^{2,+}_t]\\
   &=& \frac{1}{c_1 c_2} E_{\mathcal{Q}}[(V^{(1)}_0+c_1 V^{(1)}_L t-V^{(1)}_t +\xi_1  \int_0^t \sigma^{(1)}_s\, dW^{(1)}_s)(V^{(2)}_0+c_2 V^{(2)}_L t-V^{(2)}_t +\xi_2 \int_0^t \sigma^{(2)}_s\, dW^{(2)}_s)]\\
   &=& \frac{1}{c_1 c_2} \left[ V^{(1)}_0 V^{(2)}_0+ c_2 V^{(1)}_0  V^{(2)}_L t- V^{(1)}_0  E_{\mathcal{Q}}[V^{(2)}_t] + \xi_2 V^{(1)}_0  E_{\mathcal{Q}}[\int_0^t \sigma^{(2)}_s\, dW^{(2)}_s] \right. \\
   &+&   c_1 V^{(2)}_0 V^{(1)}_L t  +c_1 c_2 V^{(1)}_L V^{(2)}_L t^2-c_1 V^{(1)}_L t E_{\mathcal{Q}}[V^{(2)}_t]+c_1 \xi_2 V^{(1)}_L t E_{\mathcal{Q}}[ \int_0^t \sigma^{(2)}_s\, dW^{(2)}_s] \\
   &-& V^{(2)}_0 E_{\mathcal{Q}}[V^{(1)}_t]-c_2 V^{(2)}_L t E_{\mathcal{Q}}[V^{(1)}_t]+E_{\mathcal{Q}}[V^{(1)}_t V^{(2)}_t]-\xi_2 E_{\mathcal{Q}}[V^{(1)}_t \int_0^t \sigma^{(2)}_s\, dW^{(2)}_s]\\
   &+& \xi_1 V^{(2)}_0 E_{\mathcal{Q}}[ \int_0^t \sigma^{(1)}_s\, dW^{(1)}_s]+c_2 \xi_1 V^{(2)}_L t E_{\mathcal{Q}}[ \int_0^t \sigma^{(1)}_s\, dW^{(1)}_s]\\
   &-& \left. \xi_1 E_{\mathcal{Q}}[V^{(2)}_t \int_0^t \sigma^{(1)}_s\, dW^{(1)}_s]+ \xi_1 \xi_2 E_{\mathcal{Q}}[\int_0^t \sigma^{(1)}_s\, dW^{(1)}_s\int_0^t \sigma^{(2)}_s\, dW^{(2)}_s]   \right]
   \end{eqnarray*}
   Now, we have that:
   \begin{eqnarray*}
   E_{\mathcal{Q}}[ \int_0^t \sigma^{(j)}_s\, dW^{(j)}_s]  &=& 0,\; j=1,2\\
   E_{\mathcal{Q}}[\int_0^t \sigma^{(1)}_s\, dW^{(1)}_s \int_0^t \sigma^{(2)}_s\, dW^{(2)}_s] &=& E_{\mathcal{Q}}\langle \int_0^t \sigma^{(1)}_s\, dW^{(1)}_s\int_0^t \sigma^{(2)}_s\, dW^{(2)}_s \rangle= \rho_V \int_0^t m_{12}(s)\;ds\\
   E_{\mathcal{Q}}[V^{(1)}_t \int_0^t \sigma^{(2)}_s\, dW^{(2)}_s]&=& E_{\mathcal{Q}}[(V^{(1)}_0 e^{-c_1 t}+V^{(1)}_L(1-e^{-c_1 t})\\
   &+& \xi_1 e^{-c_1 t} \int_0^t e^{c_1 s}\sigma^{(1)}_s \, dW_s^{(1)})\int_0^t \sigma^{(2)}_s\, dW^{(2)}_s]\\
   &=& (V^{(1)}_0 e^{-c_1 t}+V^{(1)}_L(1-e^{-c_1 t}))E_{\mathcal{Q}}[\int_0^t \sigma^{(2)}_s\, dW^{(2)}_s]\\
   &+& \xi_1 e^{-c_1 t} E_{\mathcal{Q}}[\int_0^t e^{c_1 s}\sigma^{(1)}_s \, dW_t^{(1)}\int_0^t \sigma^{(2)}_s\, dW^{(2)}_s]\\
   &=& \xi_1 e^{-c_1 t} E_{\mathcal{Q}} \langle \int_0^t e^{c_1 s}\sigma^{(1)}_s \, dW_s^{(1)},\int_0^t \sigma^{(2)}_s\, dW^{(2)}_s \rangle\\
    &=& \xi_1 \rho_V e^{-c_1 t} \int_0^t e^{c_1 s}m_{12}(s)\;ds
      \end{eqnarray*}
   Similarly:
   \begin{equation*}
     E_{\mathcal{Q}}[V^{(2)}_t \int_0^t \sigma^{(1)}_s\, dW^{(1)}_s]=\xi_2 \rho_V e^{-c_2 t} \int_0^t e^{c_2 s}m_{12}(s)\;ds
   \end{equation*}
   Therefore:
    \begin{eqnarray*}
   mv^+_{12}(t) &:=&  \frac{1}{c_1 c_2} \left[ P_3(t)- (V^{(1)}_0+ c_1 V^{(1)}_L t) mv_{1,2}(t) - (V^{(2)}_0+c_2 V^{(2)}_L t )mv_{1,1}(t) \right.\\
   &+& ms_{12}(t) -\xi_1 \xi_2  \rho_V e^{-c_1 t} B_1(t) - \left. \xi_1 \xi_2 \rho_V e^{-c_2 t}B_2(t) + \xi_1 \xi_2  \rho_V A(t) \right]
   \end{eqnarray*}
   where:
   \begin{equation*}
     P_3(t)=V^{(1)}_0 V^{(2)}_0+ c_2 V^{(1)}_0  V^{(2)}_L t+ c_1 V^{(2)}_0 V^{(1)}_L t  +c_1 c_2 V^{(1)}_L V^{(2)}_L t^2
   \end{equation*}
   Moreover, from equation (\ref{eq:covsigmas})
    \begin{eqnarray*}
    A(t) &=& \int_0^t  ms_{12}(s) \;ds=\frac{\xi_1 \xi_2 \rho_V}{2(c_1+c_2}\left( t- \int_0^t e^{-\frac{1}{2}(c_1+c_2)s}\;ds \right)+\sigma_0^{(1)} \sigma_0^{(2)}\int_0^t e^{-\frac{1}{2}(c_1+c_2)s}\;ds\\
    &=& \frac{\xi_1 \xi_2 \rho_V}{2(c_1+c_2)}\left( t- \frac{2}{c_1+c_2}(1-e^{-\frac{1}{2}(c_1+c_2)t}) \right)+\frac{2 \sigma_0^{(1)} \sigma_0^{(2)}}{c_1+c_2}(1-e^{-\frac{1}{2}(c_1+c_2)t})\\
    B_j(t) &=& \int_0^t e^{c_j s}ms_{12}(s)\;ds = \frac{\xi_1 \xi_2 \rho_V}{2(c_1+c_2)}\left(\frac{1}{c_j}(e^{c_j t}-1) - \int_0^t e^{c_j-\frac{1}{2}(c_1+c_2)s}\;ds \right)\\
    &+& \sigma_0^{(1)} \sigma_0^{(2)}\int_0^t e^{c_j-\frac{1}{2}(c_1+c_2)s}\;ds\\
     &=& \frac{\xi_1 \xi_2 \rho_V}{2(c_1+c_2)}\left(\frac{1}{c_j}(e^{c_j t}-1) - \frac{2(-1)^j}{c_2-c_1}(e^{\frac{1}{2}(-1)^j(c_2-c_1)t}-1) \right)\\
     &+& \sigma_0^{(1)} \sigma_0^{(2)}\frac{2(-1)^j}{c_2-c_1}(e^{\frac{1}{2}(-1)^j(c_2-c_1)t}-1)
   \end{eqnarray*}
     \end{proof}
     \subsection{Appendix B: Derivatives of the Margrabe price}
     Derivatives of the Margrabe price are computed by elementary differentiation. Indeed, for the function:
\begin{equation*}
  M_4(x) =  x_{1}\,x_{2}-2\,\sqrt{x_{1}}\,\sqrt{x_{2}}\,x_{3}
\end{equation*}
We see that:
\begin{eqnarray*}
  \frac{\partial M_4(x)}{\partial x_1} &=& x_{2}-\frac{\sqrt{x_{2}}\,x_{3}}{\sqrt{x_{1}}},\; \frac{\partial M_4(x)}{\partial x_2}= x_{1}-\frac{\sqrt{x_{1}}\,x_{3}}{\sqrt{x_{2}}} \\
  \frac{\partial M_4(x)}{\partial x_3} &=& -2\,\sqrt{x_{1}}\,\sqrt{x_{2}}
\end{eqnarray*}
The second derivatives of $M_4(x)$ are:
\begin{eqnarray*}
  \frac{\partial^2 M_4(x)}{\partial x^2_1}   &=& \frac{\sqrt{x_{2}}\,x_{3}}{2\,{x_{1}}^{3/2}}, \;
 \frac{\partial^2 M_4(x)}{\partial x_1 \partial x_2}  = 1-\frac{x_{3}}{2\,\sqrt{x_{1}}\,\sqrt{x_{2}}} \\
 \frac{\partial^2 M_4(x)}{\partial x_1 \partial x_3}  &=& -\frac{\sqrt{x_{2}}}{\sqrt{x_{1}}},\;
 \frac{\partial^2 M_4(x)}{\partial^2 x_2}=\frac{\sqrt{x_{1}}\,x_{3}}{2\,{x_{2}}^{3/2}}\\
\frac{\partial^2 M_4(x)}{\partial x_2 \partial x_3}   &=& -\frac{\sqrt{x_{1}}}{\sqrt{x_{2}}},\;\frac{\partial^2 M_4(x)}{\partial x^2_3}  = 0
\end{eqnarray*}
Regarding the function:
\begin{eqnarray*}
  d_1(x) &=& M_3 M^{-\frac{1}{2}}_4(x)-\frac{1}{2}M^{\frac{1}{2}}_4(x)\\
  &=&  \frac{M_{3}}{\sqrt{x_{1}\,x_{2}-2\,\sqrt{x_{1}}\,\sqrt{x_{2}}\,x_{3}}}-\frac{\sqrt{x_{1}\,x_{2}-2\,\sqrt{x_{1}}\,\sqrt{x_{2}}\,x_{3}}}{2}
\end{eqnarray*}
where  $M_3=\log \left(\frac{ S^{(1)}_t}{ S^{(2)}_t} \right)$, the first and second derivatives of $ d_1(x)$ are:
\begin{eqnarray*}
   \frac{\partial d_1(x)}{\partial x_j} &=&- \frac{1}{2} M_3 M^{-\frac{3}{2}}_4(x)\frac{\partial M_4(x)}{\partial x_j}-\frac{1}{4}M^{\frac{1}{2}}_4(x)\frac{\partial M_4(x)}{\partial x_j}, j=1,2,3\\
    \frac{\partial^2 d_1(x)}{\partial x_j \partial x_k}  &=&   \frac{3}{4} M_3 M^{-\frac{5}{2}}_4(x)\frac{\partial M_4(x)}{\partial x_j}\frac{\partial M_4(x)}{\partial x_4}- \frac{1}{2} M_3 M^{-\frac{3}{2}}_4(x)\frac{\partial^2 M_4(x)}{\partial x_j \partial x_k}\\
    &+& \frac{1}{8}  M^{-\frac{3}{2}}_4(x)\frac{\partial M_4(x)}{\partial x_j}\frac{\partial M_4(x)}{\partial x_k}- \frac{1}{4}M^{-\frac{1}{2}}_4(x)\frac{\partial^2 M_4(x)}{\partial x_j \partial x_k}, j,k=1,2,3
\end{eqnarray*}
Finally:
\begin{eqnarray*}
   \frac{\partial C_M(x)}{\partial x_j} &=& M_1 f_Z(d_1(x))\frac{\partial d_1(x)}{\partial x_j}-M_2 f_Z(d_1(x))\frac{\partial d_1(x)}{\partial x_j}, j=1,2,3\\
   \frac{\partial^2 C_M(x)}{\partial x_j \partial x_k}  &=& M_1 \left(\frac{\partial f_Z(d_1(x))}{\partial x_k} \frac{\partial d_1(x)}{\partial x_j}+ f_Z(d_1(x))\frac{\partial^2 d_1(x)}{\partial x_j \partial x_k}  \right)\\
   &-& M_2  \left(\frac{\partial f_Z(d_2(x))}{\partial x_k} \frac{\partial d_2(x)}{\partial x_j}+ f_Z(d_2(x))\frac{\partial^2 d_2(x)}{\partial x_j \partial x_k}  \right)\\
    &=& M_1 \left(-d_1(x) f_Z(d_1(x)) \frac{\partial d_1(x)}{\partial x_j} \frac{\partial d_1(x)}{\partial x_k}+ f_Z(d_1(x))\frac{\partial^2 d_1(x)}{\partial x_j \partial x_k}  \right)\\
   &-& M_2  \left(-d_2(x) f_Z(d_2(x))\frac{\partial d_2(x)}{\partial x_j} \frac{\partial d_1(x)}{\partial x_k}+ f_Z(d_2(x))\frac{\partial^2 d_2(x)}{\partial x_j \partial x_k}  \right)
\end{eqnarray*}
  
\end{document}